\documentstyle{elsart}
\input{epsf}

\begin{document}
\unitlength = 0.0022\textwidth
\begin{frontmatter}
\title{Phase-ordering and persistence:\\
relative effects of space-discretization, chaos, and anisotropy}

\author{Julien Kockelkoren, Ana\"el Lema\^{\i}tre, and Hugues Chat\'e}
\address{CEA, Service de Physique de l'Etat Condens\'e\\
Centre d'Etudes de Saclay, 91191 Gif-sur-Yvette, France}
\date{\today}

\begin{abstract}
The peculiar phase-ordering properties of a lattice of coupled chaotic maps
studied recently (A.~Lema\^\i tre \& H.~Chat\'e, {\em Phys.~Rev.~Lett.} {\bf
82}, 1140 (1999)) are revisited with the help of detailed investigations
of interface motion. It is shown that ``normal'', curvature-driven-like
domain growth is recovered at larger scales than considered before,
and that the persistence exponent seems to be universal.
Using generalized persistence spectra, the properties of interface
motion in this deterministic, chaotic, lattice system are found to 
``interpolate'' between those of the two canonical reference
systems, the (probabilistic) Ising model, and the (deterministic), 
space-continuous, time-dependent Ginzburg-Landau equation. 
\end{abstract}
\end{frontmatter}

\section{Introduction}

Following a quench at low temperature, bistable ``ferromagnetic'' systems 
usually exhibit domain coarsening dynamics.
This phase separation process has been observed 
in various experimental setups, as well as in numerous model systems
\cite{Bray}.
The Ising model and its continuous counterpart, 
the so-called time-dependent Ginzburg-Landau (TDGL) equation 
are usually taken as paradigms
 of non-conservative coarsening dynamics \cite{MODELA}. But many 
other systems, e.g. the diffusion equation, also exhibit
phase-ordering dynamics once suitable ``phases'' are defined
(for the diffusion equation, one can for instance consider the sign of 
a zero-mean field). Spatially-extended chaotic systems such as
coupled map lattices (CMLs) can also exhibit coarsening transients, after which
one chaotic phase dominates another, leaving the system in a long-range
ordered state that is usually accompanied by a non-trivial evolution of 
spatially-averaged quantities \cite{NTCB,lc99}.

The standard theory of domain coarsening predicts
that, in general, the correlation length $L(t)$ grows  algebraically in
time, $L(t) \sim t^{1/z}$ \cite{Bray,MODELA}.
Moreover, the exponent usually takes the value $z=2$ in systems 
with a non-conserved, scalar order parameter when the domain growth is 
driven by curvature.
This universality of domain growth processes is now well established.
Another important quantifier of phase ordering dynamics is persistence 
\cite{Vishnou},
defined, e.g., as the fraction of space that has remained 
in the same phase since some given initial time $t_0$.
Persistence is seen to decay algebraically $p(t)\propto t^{-\theta}$
with exponent $\theta$ in systems with algebraic domain growth,
reflecting the stationarity of two-time correlations expressed in logarithmic
time.
As a matter of fact, the correlation length $L(t)$ provides 
a better, ``natural'' measure of time, by which persistence decays
as $p(t)\propto L(t)^{-\bar\theta}$ (with $\bar\theta=z\theta$)
even in systems with, say, logarithmic growth of domains. 
The (degree of) universality of persistence exponents is still largely an
open question today.  Even models as close to each other as the
zero-temperature Ising model and the  TDGL equation ---they
share the same exponent $z=2$, the same structure function, 
the same Fisher-Huse exponent--- seem to possess different
persistence exponents (currently available measurements give
$\theta = 0.20$ for TDGL \cite{CS,BCDL} and $0.22$ for Ising 
\cite{PER-ISING,BCDL}).

In fact, since it depends on the whole history of each point in the space,
persistence is a rather complex, non-local quantity. Exact or approximate
values of persistence exponents are usually not simple numbers. Understanding
the extent to which persistence properties are universal is one of the
challenging problems of modern statistical physics. 
It is also important from an experimental point of view, since persistence is
an easily measurable quantifier of phase ordering.
One can hope to 
bring new light to this problem by studying non-conventional models like
CMLs. In a sense, they are not constrained as are traditional models:
their local dynamics cannot be rigorously reduced to that of standard
models: there is no Hamiltonian, no detailed balance, and they have been
shown (numerically) to exhibit Ising-like phase transitions with
critical exponents that are 
significantly different from those of the Ising model \cite{Marcq}.

In~\cite{lc99}, domain growth has been studied in simple CMLs.
Numerical simulations showed the expected scaling behavior, but with 
exponents $z$ and $\theta$ continuously varying with
the strength of the diffusive coupling between chaotic units, although
$\bar{\theta}$ was found to be universal.
It was also found that ``normal'', TDGL, values were recovered in the
continuous space limit of CMLs, suggesting that a role is played by
lattice effects.

In this article, we come back to these somewhat surprising  results
and study in some detail the dynamics of the interfaces delimiting 
domains, in order to unravel the origin of the
peculiar behavior observed on more global quantities, such as $L(t)$.

\section{Domain growth and interface dynamics in chaotic CMLs}

We consider  $d=2$ dimensional square  lattices ${\cal L}$
of diffusively-coupled identical maps $S_\mu$ acting on real variables
$(X_{\vec r})_{{\vec r} \in {\cal L}}$:
\begin{equation}
X_{\vec r}^{t+1} = (1-2dg) S_{\mu}(X_{\vec r}^t) + 
g \sum_{{\vec e} \in {\cal V}} S_{\mu}(X_{{\vec r}+{\vec e}}^t) \;,
\label{eq-cml}
\end{equation}
where ${\cal V}$ is the set of $2d$ nearest neighbors $\vec e$ of site
$\vec 0$.
For simplicity, we present results obtained for the piecewise linear, odd,
local map $S_{\mu}$ defined by:
\begin{equation}
S_{\mu}(X) = \left\{
\begin{array}{lll}
\mu X & {\rm if} & X \in [-1/3,1/3] \\
2\mu/3 - \mu X & {\rm if} & X \in [1/3,1] \\
-2\mu/3 - \mu X & {\rm if} & X \in [-1,-1/3] \; .
\end{array} \right.
\label{eq-mhmap}
\end{equation}

Choosing $\mu\in[1,2]$ guarantees that the local map has two
symmetric invariant intervals, allowing us to define ``spins'' as
$\sigma_{\vec r}= {\rm sign}(X_{\vec r})$.
The deterministic system thus defined is similar to the Ising model 
at zero temperature in the sense that local variables can flip only 
when crossed by an interface.

\begin{figure}
\begin{center}
\begin{picture}(200,200)(0,0)
\put(110,210){\makebox(0,0){\large (a)}}
\put(0,0){\makebox(200,200){\epsfxsize=180\unitlength\epsffile{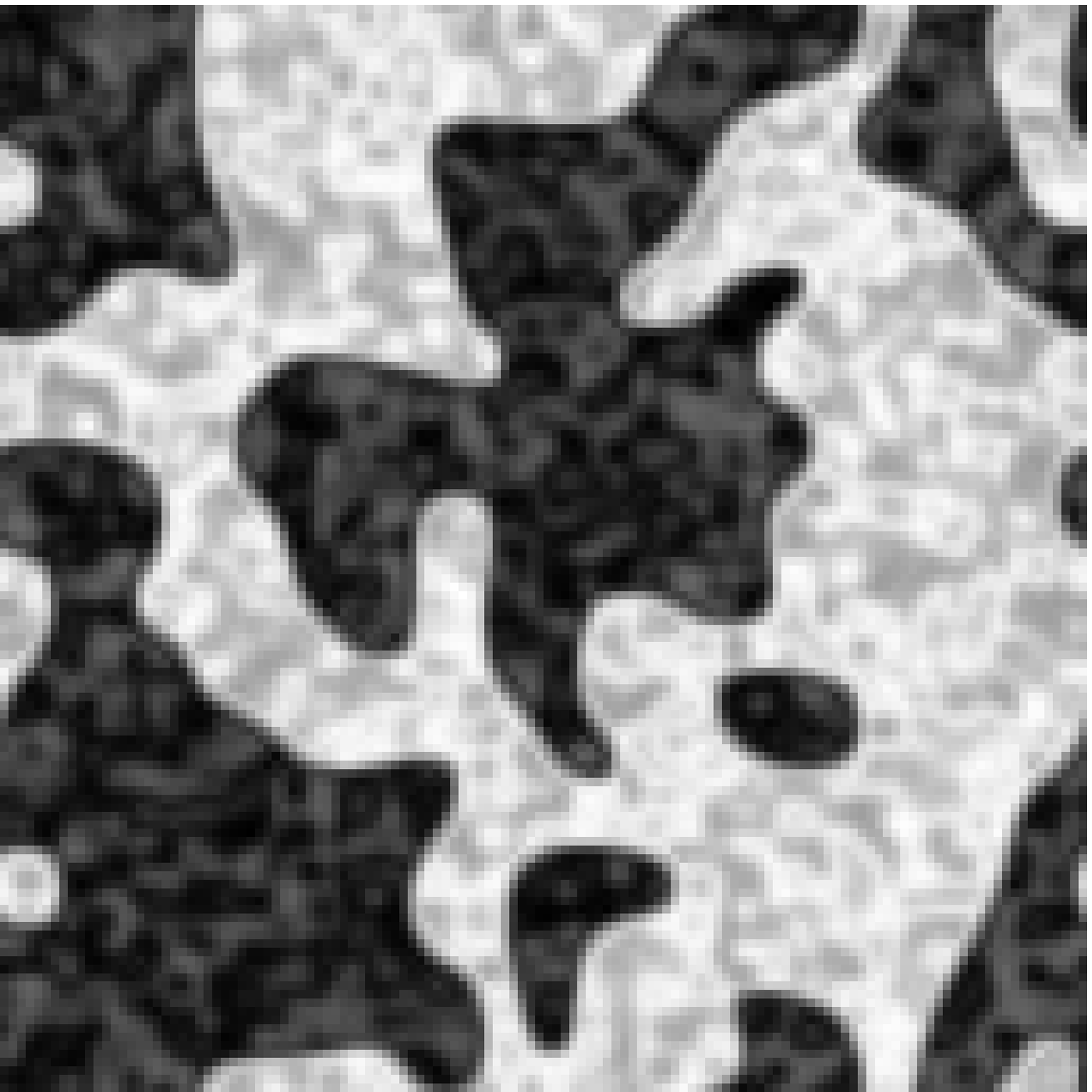}}}
\end{picture}
\hspace{30\unitlength}
\begin{picture}(200,200)(0,0)
\put(110,210){\makebox(0,0){\large (b)}}
\put(0,0){\makebox(200,200){\epsfxsize=180\unitlength\epsffile{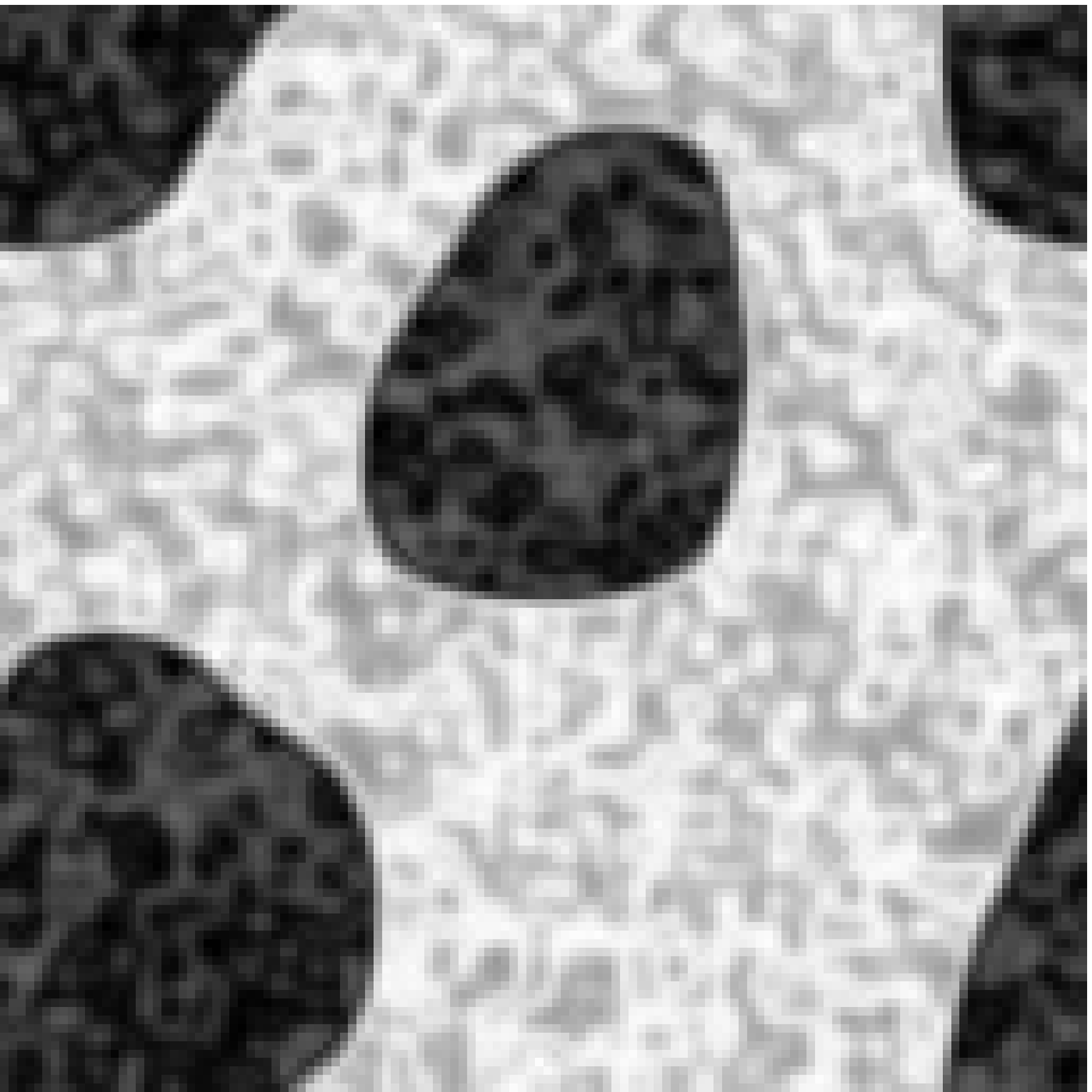}}}
\end{picture}
\end{center}
\vspace{20\unitlength}
\begin{center}
\begin{picture}(200,200)(0,0)
\put(110,210){\makebox(0,0){\large (c)}}
\put(0,0){\makebox(200,200){\epsfxsize=180\unitlength\epsffile{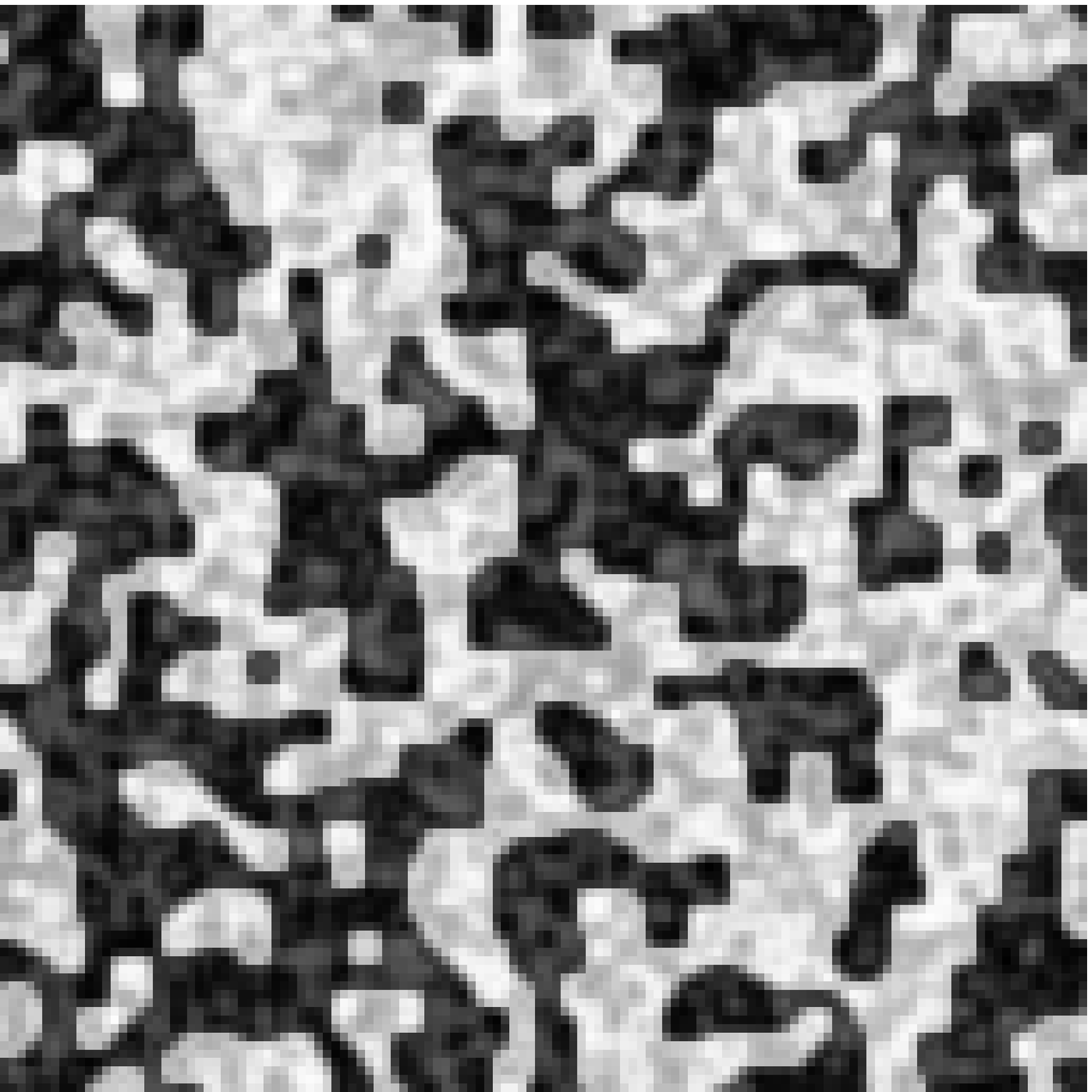}}}
\end{picture}
\hspace{30\unitlength}
\begin{picture}(200,200)(0,0)
\put(110,210){\makebox(0,0){\large (d)}}
\put(0,0){\makebox(200,200){\epsfxsize=180\unitlength\epsffile{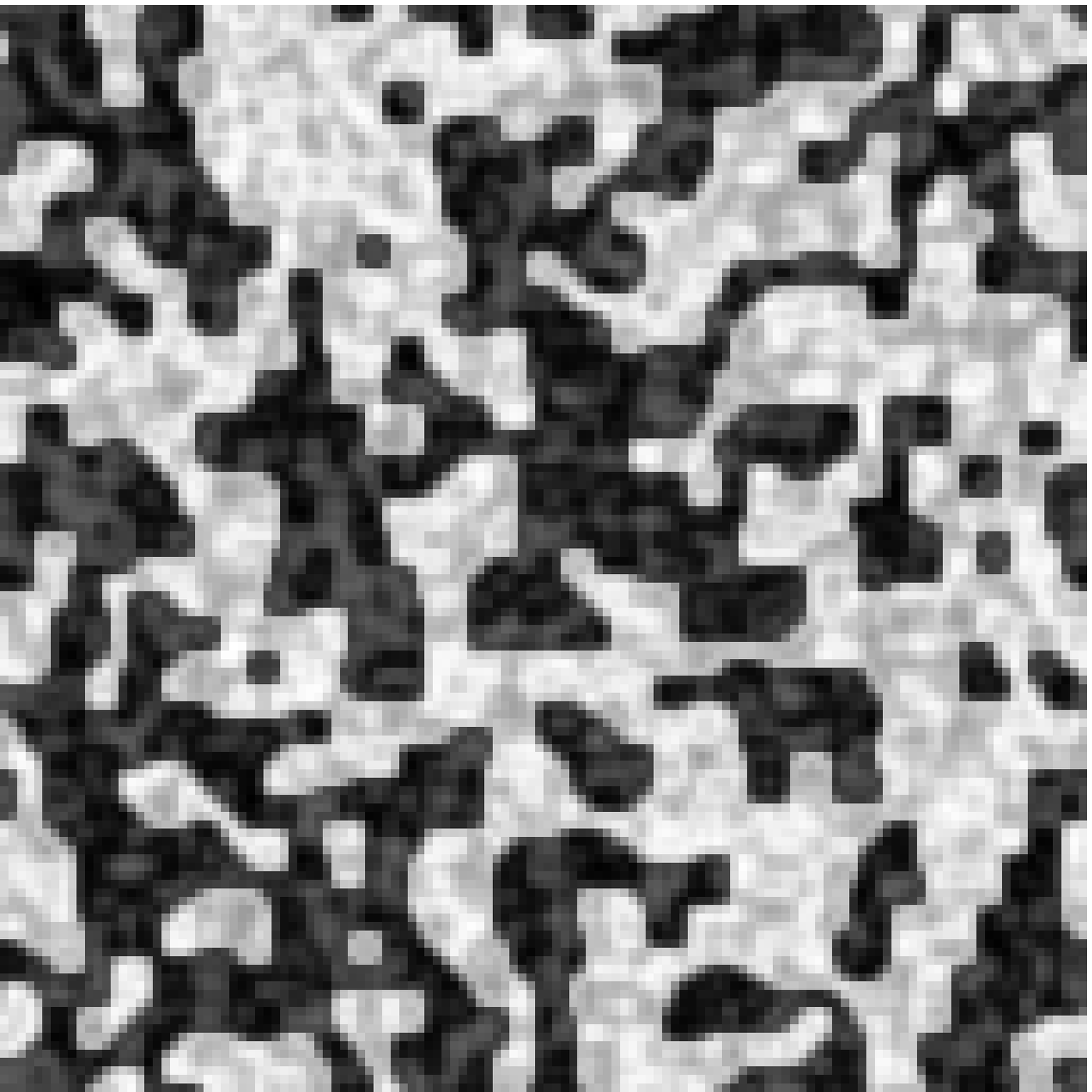}}}
\end{picture}
\end{center}
\caption{Snapshots of the $d=2$ CML with local map (\protect\ref{eq-mhmap}).
Lattice of $128^2$ sites, grey scale from $X=-1$ (white) to $X=1$ (black),
uncorrelated initial conditions.
(a,b): transient leading to complete ordering at $g=0.2>g_{\rm e}$,
$t=100$ and 1000; 
(c,d): blocked state at $g=0.15<g_{\rm e}$, $t=1000$ and 2000.}
\label{fig-snap}
\end{figure}

Take $\mu=1.9$ and consider different choices for coupling strength $g$.
Two cases are illustrated in Fig.~\ref{fig-snap}: at small $g$,
the evolution leads to blocked clusters of the two ``phases'' 
corresponding to the two invariant intervals (which ``shrink'' but 
are preserved by the linear diffusive coupling).
Domain walls are pinned between lattice sites; the system is multistable.
On the contrary, for larger values of the coupling,
domain coarsening never stops, leading to the emergence 
of long-range order. 

Thus, there exists a threshold value $g_{\rm e}$ separating these
regimes: domain growth is expected to slow down as $g$ goes to $g_{\rm e}$
from above, corresponding to a gradual ``freezing'' of interface dynamics.

\subsection{Previous results}

We first review 
some of the results presented in \cite{lc99}
for the sake of completeness. 

\begin{figure}
\begin{center}
\begin{picture}(200,200)(0,0)
\put(0,0){\makebox(200,200){\epsfxsize=220\unitlength\epsffile{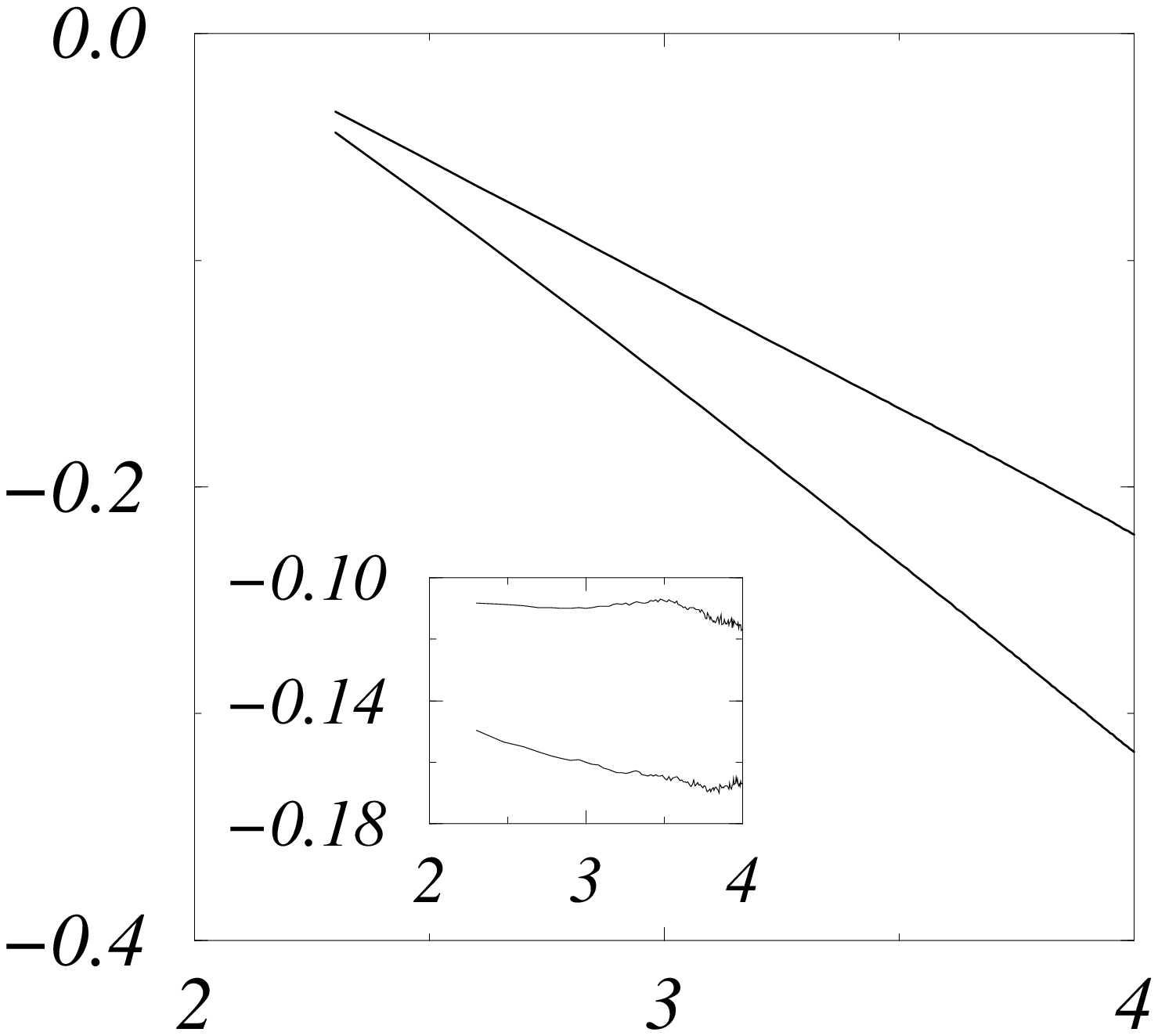}}}
\put(110,210){\makebox(0,0){(a)}}
\put(0,170){\makebox(0,0){\large $\log p$}}
\put(180,0){\makebox(0,0){\large $\log t$}}
\put(170,150){\makebox(0,0){$\scriptstyle g=0.172$}}
\put(170,140){\vector(0,-1){20}}
\put(170,50){\makebox(0,0){$\scriptstyle g=0.185$}}
\put(150,60){\vector(1,1){20}}
\end{picture}
\hspace{30\unitlength}
\begin{picture}(200,200)(0,0)
\put(0,0){\makebox(200,200){\epsfxsize=220\unitlength\epsffile{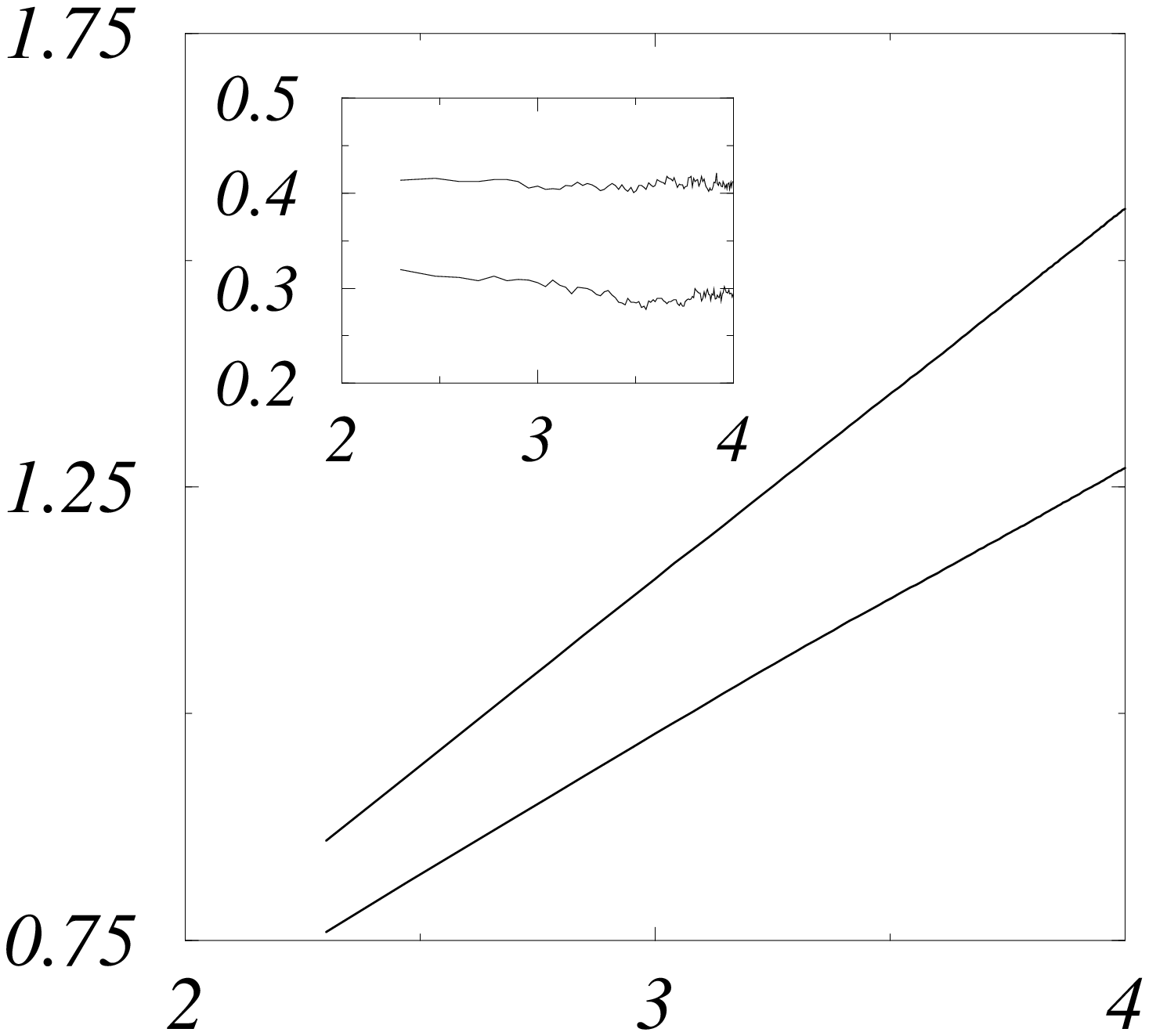}}}
\put(110,210){\makebox(0,0){(b)}}
\put(0,170){\makebox(0,0){\large $\log L$}}
\put(180,0){\makebox(0,0){\large $\log t$}}
\put(60,90){\makebox(0,0){$\scriptstyle g=0.185$}}
\put(45,80){\vector(1,-1){20}}
\put(170,50){\makebox(0,0){$\scriptstyle g=0.172$}}
\put(180,60){\vector(-1,1){20}}
\end{picture}
\end{center}
\vspace{20\unitlength}
\begin{center}
\begin{picture}(200,200)(0,0)
\put(0,0){\makebox(200,200){\epsfxsize=220\unitlength\epsffile{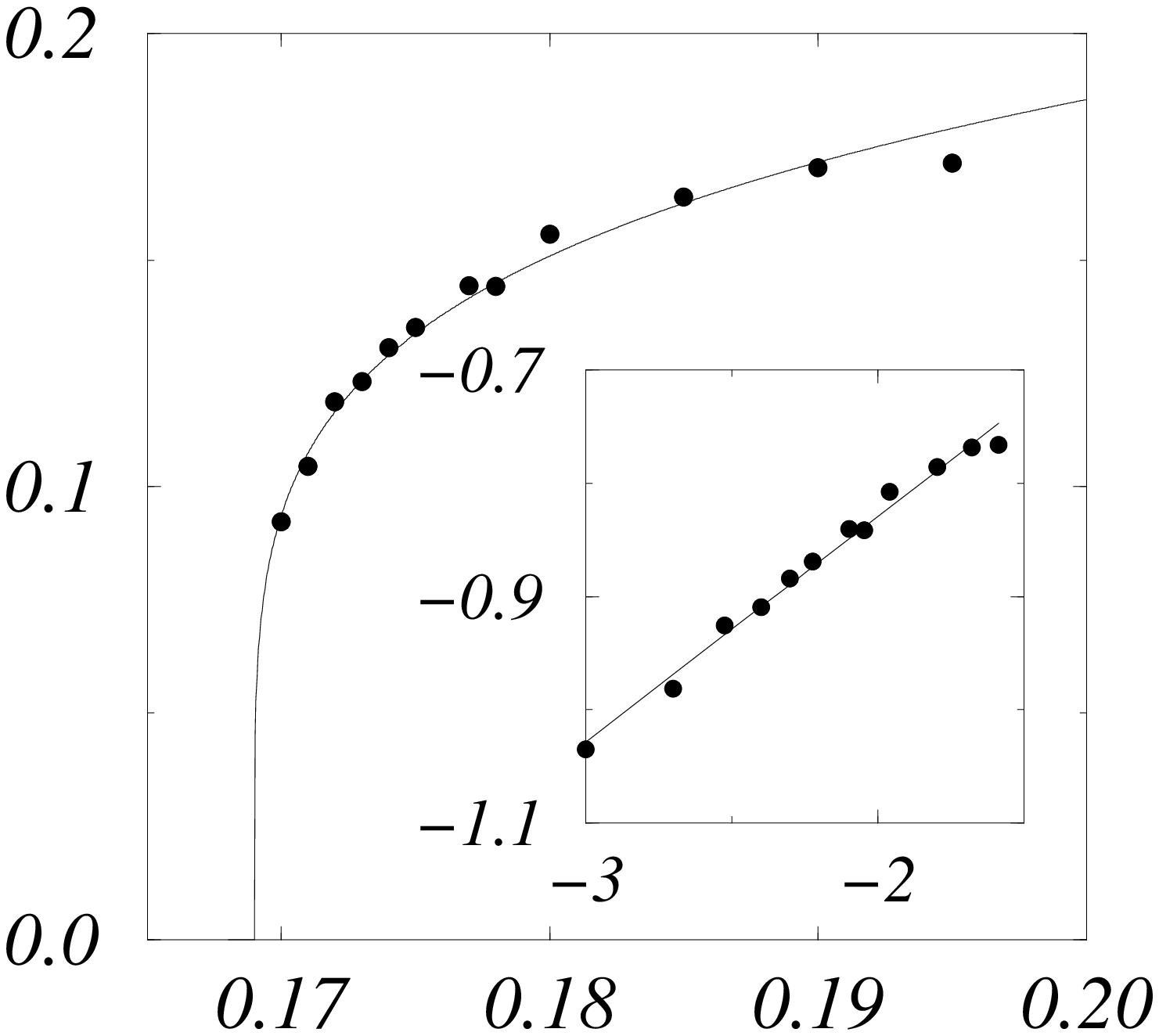}}}
\put(110,210){\makebox(0,0){(c)}}
\put(0,165){\makebox(0,0){\large $\theta$}}
\put(180,0){\makebox(0,0){\large $g$}}
\put(125,115){\makebox(0,0){$\log \theta$}}
\put(155,60){\makebox(0,0){$\log(g\!-\!g_{\rm e}\!)$}}
\end{picture}
\hspace{30\unitlength}
\begin{picture}(200,200)(0,0)
\put(0,0){\makebox(200,200){\epsfxsize=220\unitlength\epsffile{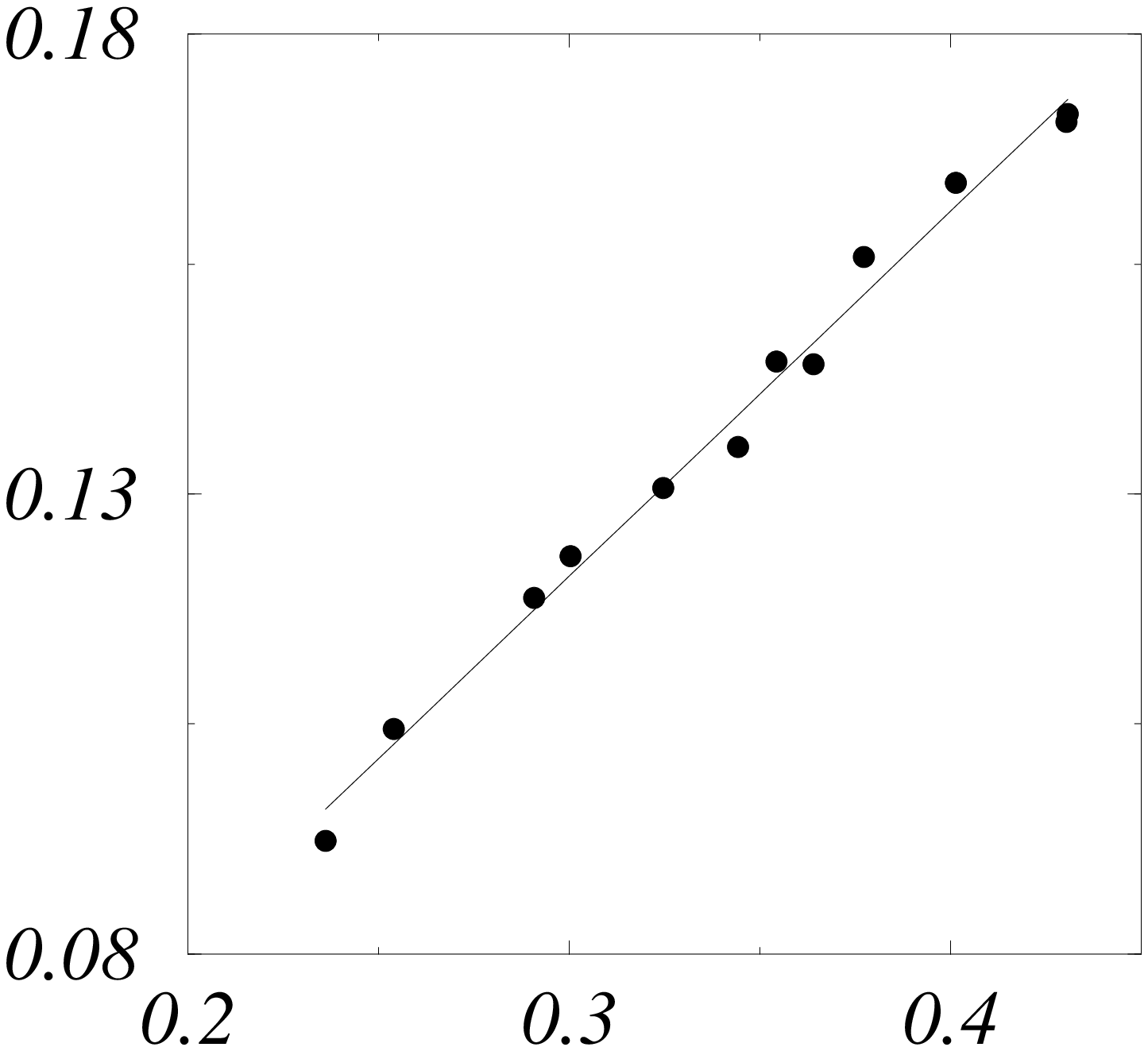}}}
\put(110,210){\makebox(0,0){(d)}}
\put(0,165){\makebox(0,0){\large $\theta$}}
\put(190,0){\makebox(0,0){\large $1/z$}}
\end{picture}
\end{center}
\caption{Domain coarsening on a lattice of $2048^2$ sites, up to time $t=10^4$.
(a): log-log plot of $p(t)$ vs time for two values of the coupling strength
($g=0.172,0.185$); local slope in insert
(b): $L(t)$ in log-log for the same runs with local slope in insert
(c): exponent $\theta$ as a function of the coupling strength $g$,
measured from plots similar to (a).
(d): linear dependence of persistence exponent $\theta$ with $1/z$.
}
\label{f2}
\end{figure}

Starting from random initial conditions with exactly zero magnetization
(for the ``spin'' variables $\sigma_{\vec r}= {\rm sign}(X_{\vec r})$), 
we measured $L(t)$ defined by the two-point correlation function at mid-height,
as well as $p(t)$, the persistent fraction of sites since $t_0=0$.
Fig.~\ref{f2} shows the results obtained from single runs 
on lattices of $2048^2$ sites
simulated up to time $t=10^4$ (at later times, finite-size fluctuations
become too important). Algebraic growth of $L(t)$ and decay of $p(t)$ 
is observed, but with effective exponents $z$ and $\theta$ which vary
with $g$, the coupling strength. The decay of $\theta$ with $g$ is 
best fitted with some power-law dependence which defines the 
threshold coupling $g_{\rm e}\simeq 0.169(1)$.

Normal, $z=2$ coarsening is only recovered when one approaches the 
continuous-space limit of CMLs which, in practice, consists in applying
the coupling operator many times. In this limit, the persistence exponent
$\theta$ approaches the value known for the TDGL equation, i.e.,
$\theta \simeq 0.20$.
The underlying lattice thus seems to influence the
propagation of fronts in the system, by discretization and/or 
anisotropy effects.

\subsection{Interface dynamics}
\label{interface}

Anisotropic curvature-driven domain growth can usually be
defined by the following expression for the normal velocity of the interface 
\cite{spohn}:
\begin{equation}
v_{\rm n} = M(\varphi) [ \sigma(\varphi) + \sigma''(\varphi)) \kappa +h ]
\label{normal-velocity} 
\end{equation}
where $M(\varphi)$ is the mobility, $\sigma(\varphi)$ the surface tension,
$\kappa$ the local curvature, $h$ an ``external field'',
and $\varphi$ an angle specifying the orientation
of the vector normal to the interface in space.

It is thus important to know whether these quantities can be defined 
in the CMLs of interest here, and, in the affirmative case, to try to
estimate them in order to investigate possible discrepancies with
Eq.(\ref{normal-velocity}).

\begin{figure}
\begin{center}
\begin{picture}(200,200)(0,0)
\put(0,0){\makebox(200,200){\epsfxsize=220\unitlength\epsffile{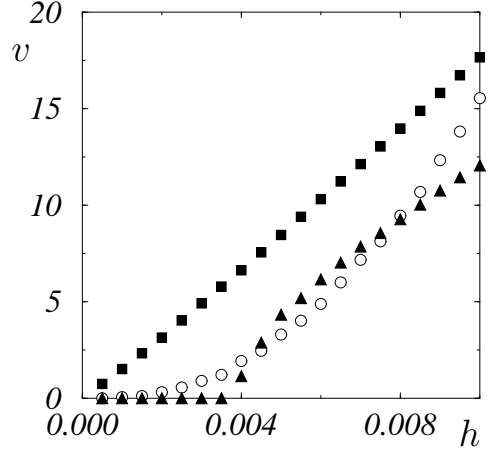}}}
\put(0,175){\makebox(0,0){\large $v$}}
\put(195,10){\makebox(0,0){\large $h$}}
\end{picture}
\hspace{30\unitlength}
\end{center}
\caption{Normal velocity  $v_{\rm n}$ of inclined straight 
interfaces under the ``external field'' $h$
for orientations $\varphi=0$ (triangles), $\varphi = \frac14 \pi$ (circles)
and $\varphi=\arctan(3/8)$ (squares) (for $g=0.2$, lattices of linear size
$\simeq 500$, from runs of 1000 iterations).}
\label{f3}
\end{figure}

The mobility is usually measured as the response of a flat interface 
($\kappa=0$) to the
driving field $h$.  
Previous work has indicated that the mobility may
not be defined in deterministic lattice systems \cite{oppo,mallet}. 
This is also the case
of the CMLs studied here. Replacing the evolution equation
(\ref{eq-cml}) by:
\begin{equation}
X_{\vec r}^{t+1} = h+ (1-h)((1-2dg) S_{\mu}(X_{\vec r}^t) + 
g \sum_{{\vec e} \in {\cal V}} S_{\mu}(X_{{\vec r}+{\vec e}}^t)) \;,
\label{eq-cml-h}
\end{equation}
one observes moving interfaces for large-enough $h$, but the velocity may 
not be constant, and the $h$ dependence of  the average velocity 
is not always linear. Moreover,
there generally exists a finite critical field value below which the
interface does not move (Fig.~\ref{f3}). In fact, the mobility seems 
to be ill-defined for all rational values of $\tan\varphi$. 
Thus the mobility of the interfaces of the CML studied here is not 
well defined, a further indication of the special character of domain
growth.

\begin{figure}
\begin{center}
\begin{picture}(200,200)(0,0)
\put(110,210){\makebox(0,0){\large (a)}}
\put(0,175){\makebox(0,0){\large $V$}}
\put(200,0){\makebox(0,0){\large $t$}}
\put(70,90){\makebox(0,0){$\pi r_d^2$}}
\put(80,90){\vector(2,1){20}}
\put(180,90){\makebox(0,0){$\pi r_t^2$}}
\put(170,90){\vector(-2,-1){20}}
\put(0,0){\makebox(200,200){\epsfxsize=220\unitlength\epsffile{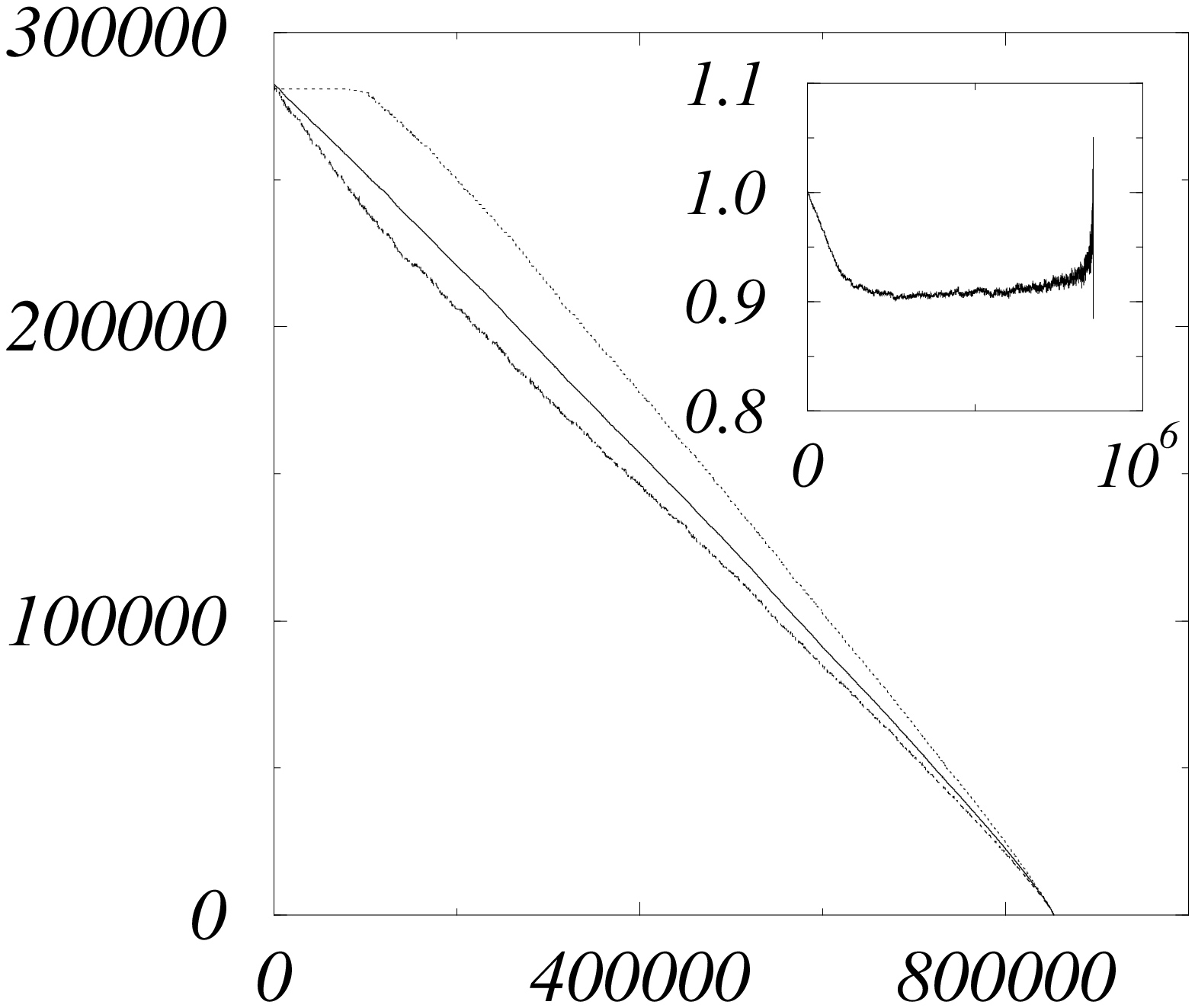}}}
\end{picture}
\hspace{30\unitlength}
\begin{picture}(200,200)(0,0)
\put(0,175){\makebox(0,0){\large $\log T_r$}}
\put(175,0){\makebox(0,0){\large $\log V$}}
\put(110,210){\makebox(0,0){\large (b)}}
\put(110,170){\makebox(0,0){$\alpha=1.03$}}
\put(60,130){\makebox(0,0){$\alpha=1.15$}}
\put(0,0){\makebox(200,200){\epsfxsize=220\unitlength\epsffile{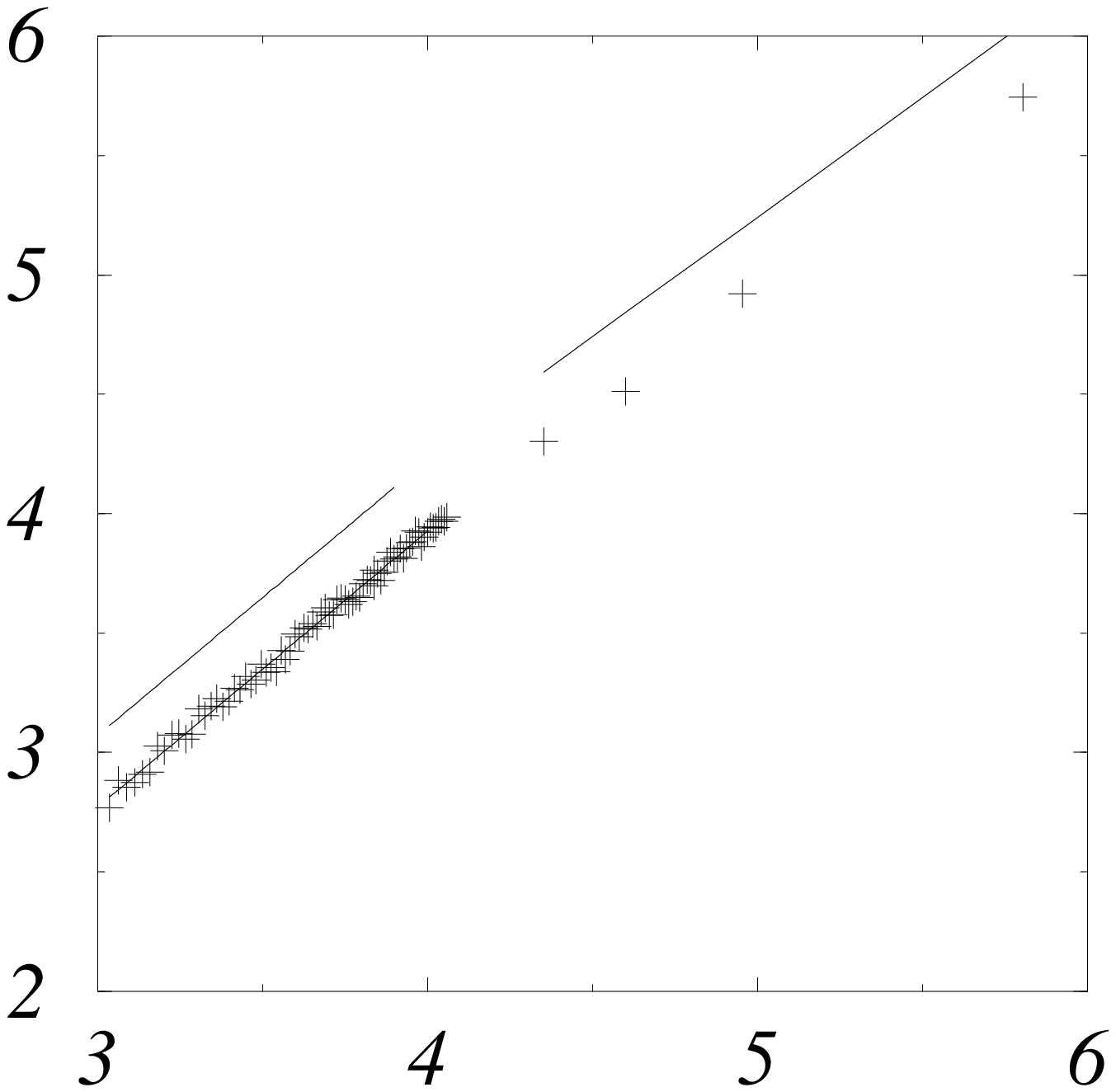}}}
\end{picture}
\end{center}
\vspace{40\unitlength}
\begin{center}
\begin{picture}(200,200)(0,0)
\put(0,175){\makebox(0,0){\large $\log T$}}
\put(180,0){\makebox(0,0){\large $\log V$}}
\put(110,210){\makebox(0,0){\large (c)}}
\put(135,180){\makebox(0,0){$\alpha=1.09$}}
\put(110,160){\makebox(0,0){$\alpha=1.19$}}
\put(0,0){\makebox(200,200){\epsfxsize=220\unitlength\epsffile{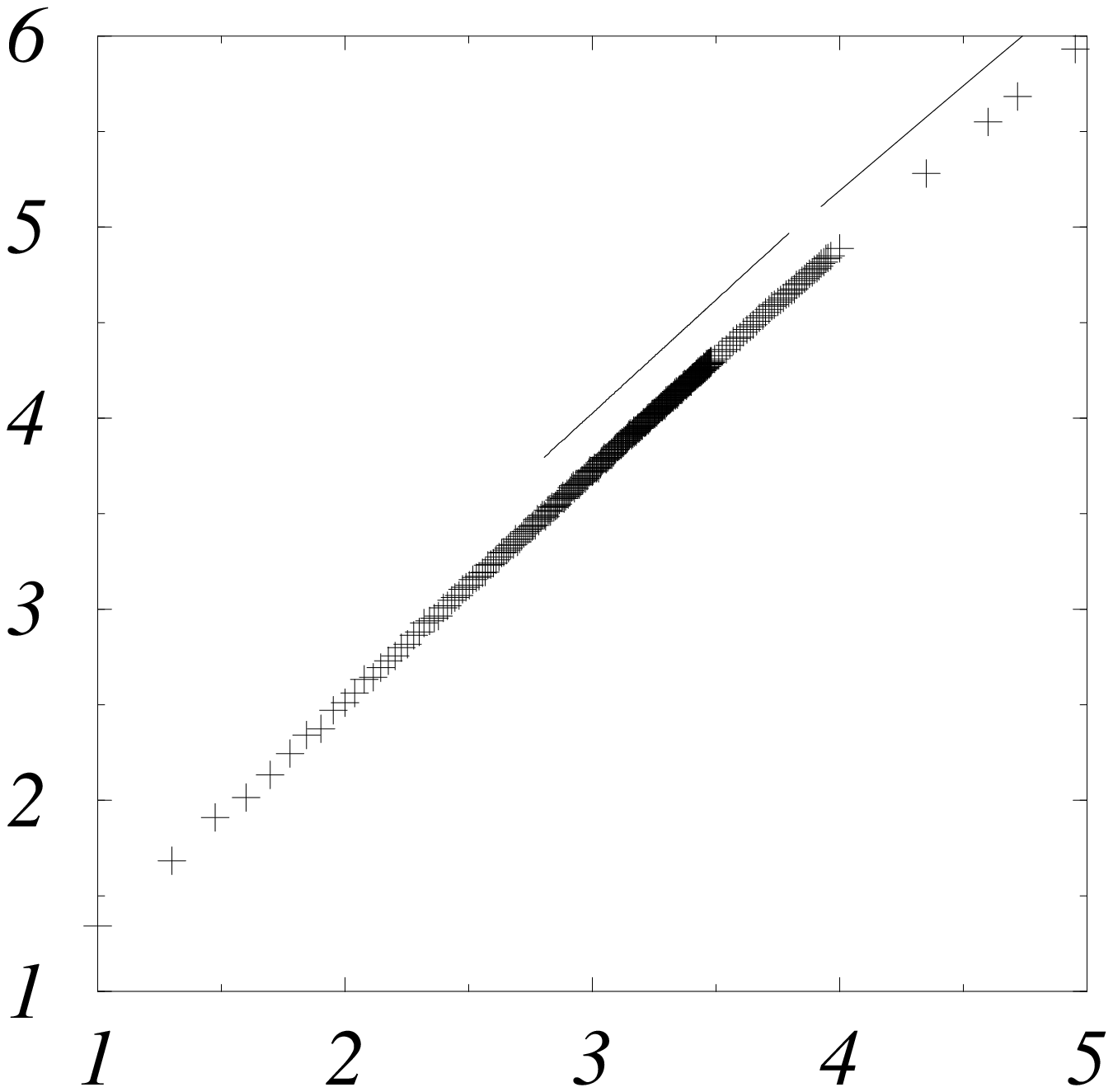}}}
\end{picture}
\hspace{30\unitlength}
\begin{picture}(200,200)(0,0)
\put(0,175){\makebox(0,0){\large $V/V(0)$}}
\put(175,0){\makebox(0,0){\large $t/T$}}
\put(165,180){\makebox(0,0){$\scriptstyle r_i=14$}}
\put(165,168){\makebox(0,0){$\scriptstyle r_i=100$}}
\put(165,156){\makebox(0,0){$\scriptstyle r_i=300$}}
\put(110,210){\makebox(0,0){\large (d)}}
\put(0,0){\makebox(200,200){\epsfxsize=220\unitlength\epsffile{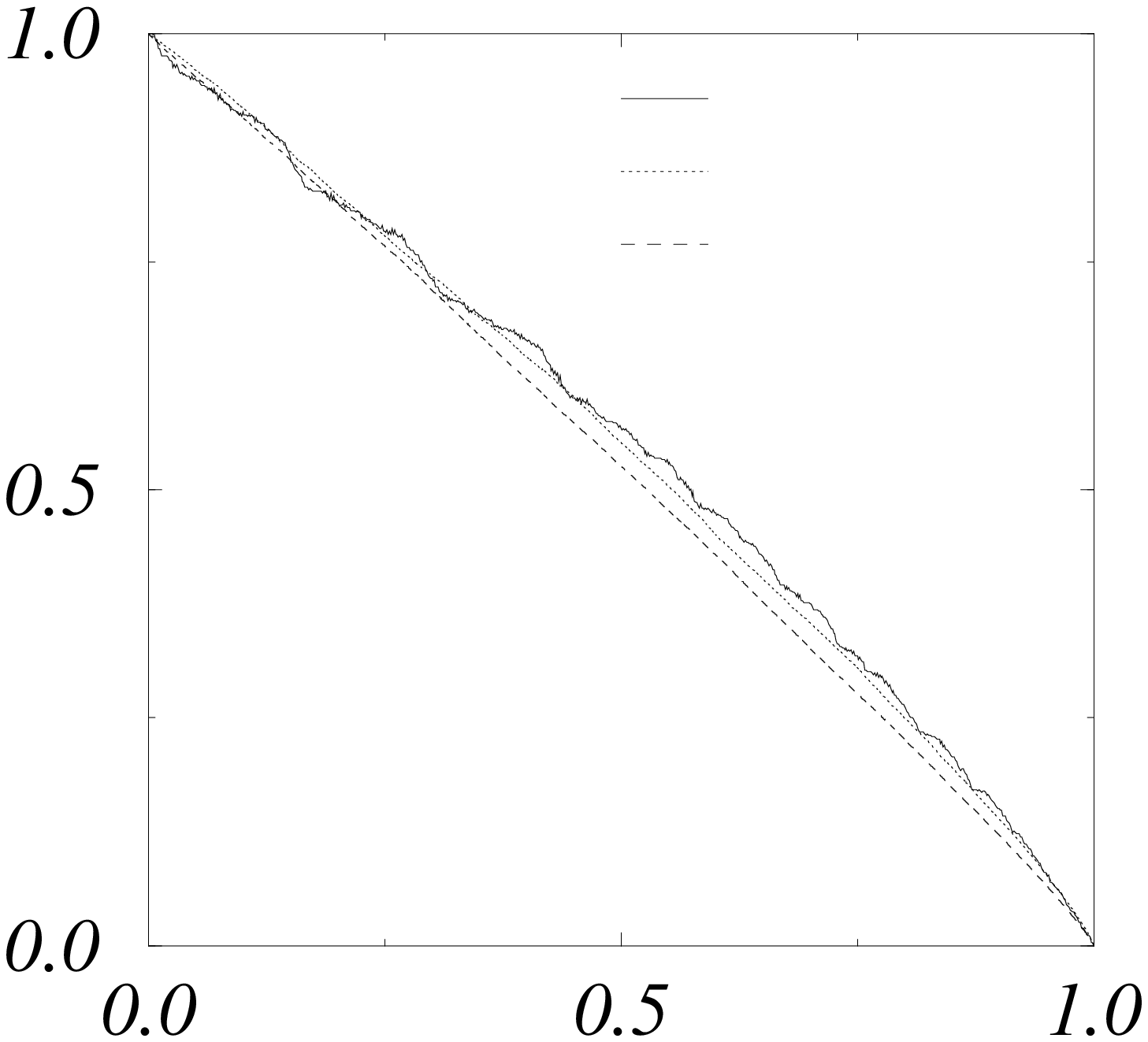}}}
\end{picture}
\end{center}
\caption{Droplet evolution. (a) Time evolution of a circular droplet:
volume $V$, $\pi r_t^2$, and  $\pi r_d^2$ vs time, where $r_t$ and $r_d$ are
radii along, respectively, lattice and diagonal directions.
Inset: ratio $r_d/r_t$ of the two radii. Radius $r_t$
remains initially constant during a transient ``rearrangement'' time.  
(b) Scaling of the rearrangement time $T_r$ with initial volume of the 
droplet. The fit for ``small''
values of $V$ (up to $10^4$) yields an exponent $\alpha=1.15$, while
``normal'' scaling is recovered  for large values ($\alpha\simeq 1$) 
(c) Scaling of the lifetime $T$ with initial volume. The fit for $V$
 yields $\alpha=1.19$, while $\alpha=1.09$, not far from 1, for large
values of $V$. 
(d) For different initial volumes: rescaled volumes $V(t)/V(0)$ as a
function of rescaled time
$t/T$. The larger the initial radius, the straighter the curves.}
\label{f4}
\end{figure}

Another typical experiment to try to assess the validity of
Eq.~(\ref{normal-velocity}) is to study the fate of droplets of
one phase immersed in the other phase. In anisotropic situations such as
those described by Eq.~(\ref{normal-velocity}),
one expects, from some circular initial droplet, a transient stage
during which the droplet takes its asymptotic shape and then shrinks 
self-similarly, its volume decreasing linearly with time;
its lifetime is therefore proportional to its initial volume \cite{langer}.
This is roughly what is observed with the CML studied here,
except that the expected scaling is recorded only for droplets
of large initial radius  (Fig.\ref{f4}). For smaller size, one
finds that the lifetime of droplets scales like $r^{2\alpha}$,
with $1/2\alpha \simeq 1/z$, where $z$ is the growth exponent
measured in Fig.\ref{f2}.
Interestingly, the crossover size is approximately of the order of
the typical size of domains at the end of the runs shown in Fig.\ref{f2}.

This suggests that the non-trivial scalings laws reported in  Fig.\ref{f2}
might only be transient, ultimately leaving place to normal 
(curvature-driven-like) phase ordering. 

\subsection{Large-scale domain growth}

\begin{figure}
\begin{center}
\begin{picture}(200,200)(0,0)
\put(110,210){\makebox(0,0){\large (a)}}
\put(0,0){\makebox(200,200){\epsfxsize=220\unitlength\epsffile{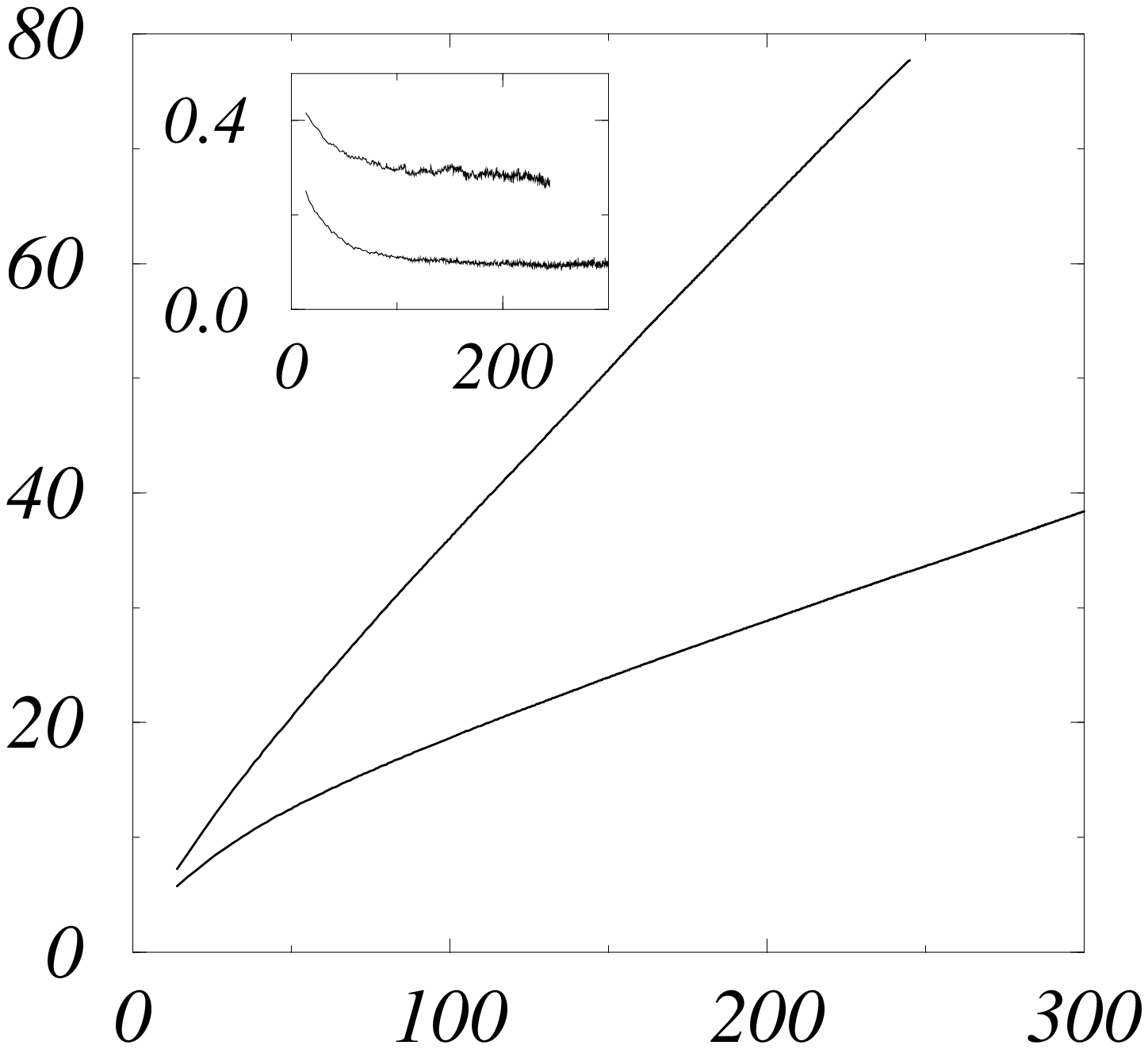}}}
\put(0,175){\makebox(0,0){\large $L$}}
\put(175,0){\makebox(0,0){\large $\sqrt{t}$}}
\put(60,115){\makebox(0,0){$\scriptstyle g=0.185$}}
\put(45,105){\vector(1,-1){20}}
\put(170,50){\makebox(0,0){$\scriptstyle g=0.172$}}
\put(180,60){\vector(-1,1){20}}
\end{picture}
\hspace{30\unitlength}
\begin{picture}(200,200)(0,0)
\put(110,210){\makebox(0,0){\large (b)}}
\put(0,0){\makebox(200,200){\epsfxsize=220\unitlength\epsffile{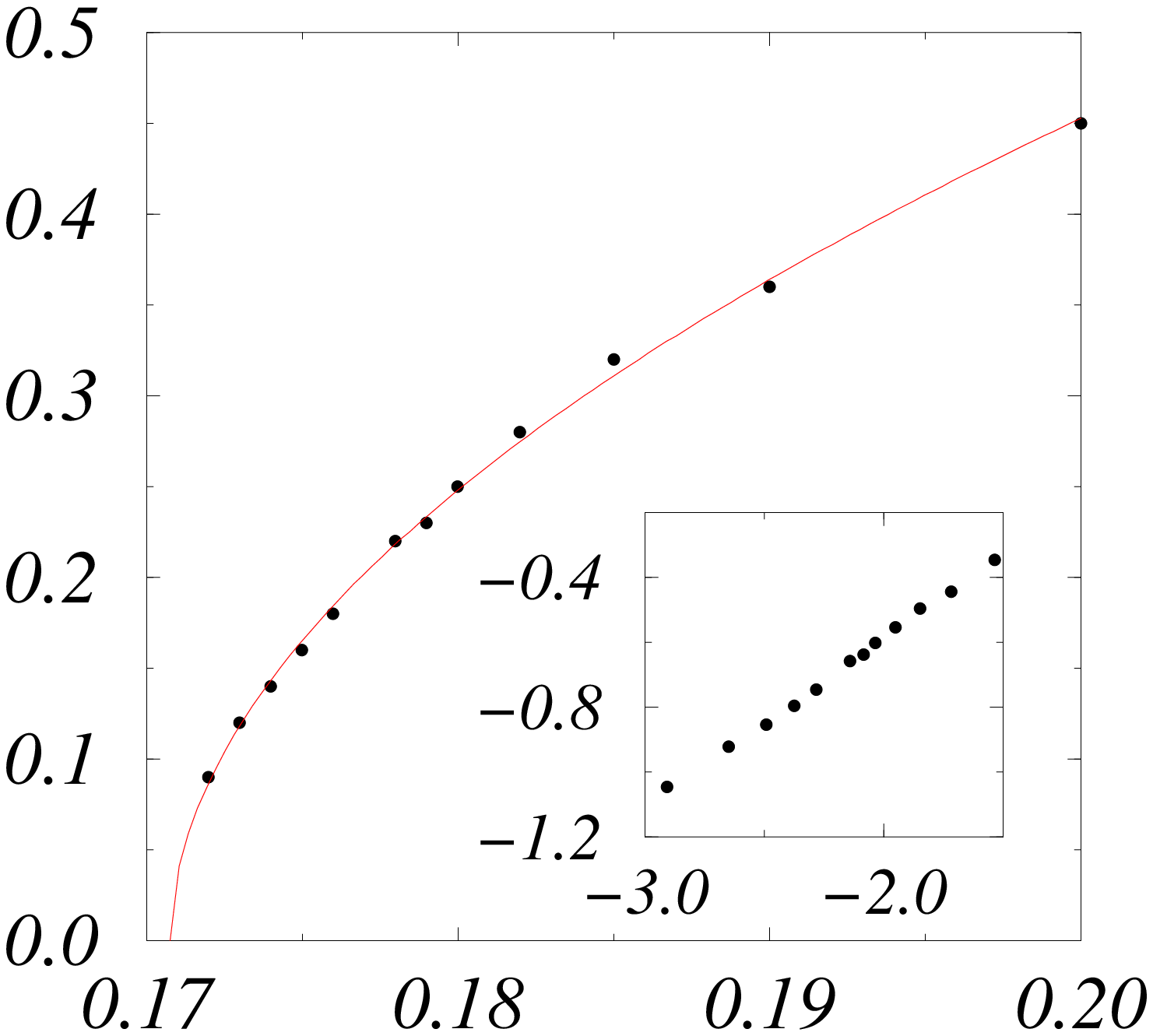}}}
\put(0,175){\makebox(0,0){\large $A$}}
\put(175,0){\makebox(0,0){\large $g$}}
\end{picture}
\hspace{30\unitlength}
\end{center}
\caption{
Domain coarsening for our CML on a lattice of $8192^2$ sites for two values of 
the coupling strength ($g=0.172, 0.185$):
(a) $L(t)$ vs. $\sqrt{t}$: 
normal coarsening is recovered after a transient of order $10^4$ timesteps,
so that, in the asymptotic regime: 
$L(t)\propto A(g)\sqrt{t}$.
(b) Effective mobility $A(g)$ vs. $g$ from simulations 
of lattices of linear size at least $4096^2$ iterated during $10^5$ timesteps. 
The fit $A(g) \simeq (g-g_{\rm e})^{1/2}$ allows one
to estimate the threshold $g_{\rm e}\simeq 0.171$ 
below which domain wall motion freezes.
Insert: $\log A(g)$ vs $\log (g-g_{\rm e})$.
}
\label{f5}
\end{figure}

We thus went back to domain growth transients following random initial
conditions, but with larger system sizes and longer simulation times than
before. Fig.~\ref{f5} shows the growth of $L(t)$ for a system of $8192^2$ sites
during $10^5$ timesteps (runs performed on a parallel machine for a total
of approximately 2000 cpu hours). 
A log-log plot of $L(t)$ reveals an increase of the local slope at long
times, which, however, does not reach the ``normal'' value
$1/z=1/2$. On the other hand, 
the emergence of the normal growth behavior is clearly 
seen in a plot of $L$ vs $t^{1/2}$, which becomes linear after 
approximately $10^4$ iterations (Fig.~\ref{f5}a).
The approach of the freezing transition at $g=g_{\rm e}$ is therefore
contained in the prefactor of the growth law $L(t) \propto A(g) t^{1/2}$.
The effective ``mobility'' $A(g)$ seems to decrease continuously to zero.
An excellent fit of the data is
$A(g) \simeq (g- g_{\rm e})^w$ with $w=0.52(3)$ 
(Fig.~\ref{f5}b), 
yielding the new estimate $g_{\rm e} \simeq 0.171(1)$, but we cannot exclude
a linear behavior near threshold which would yield a $g_{\rm e}$ value
closer to our previous estimate in \cite{lc99}.

To sum up: domain coarsening in the CML studied here does show
discrepancies with Eq.~(\ref{normal-velocity}) (such as the ill-defined
character of the mobility), but this seems to  quantitatively  influence
the dynamics at early times  only, so that normal growth behavior is recovered
asymptotically at long times. It must be noted, though, that the effect
of these features, specific to lattice determinitic systems, decreases
rather slowly in time, as testified by the long crossover times recorded
in our simulations.

\section{Persistence issues}
\subsection{Persistence at late times}

\begin{figure}
\begin{center}
\begin{picture}(200,200)(0,0)
\put(110,210){\makebox(0,0){\large (a)}}
\put(0,0){\makebox(200,200){\epsfxsize=220\unitlength\epsffile{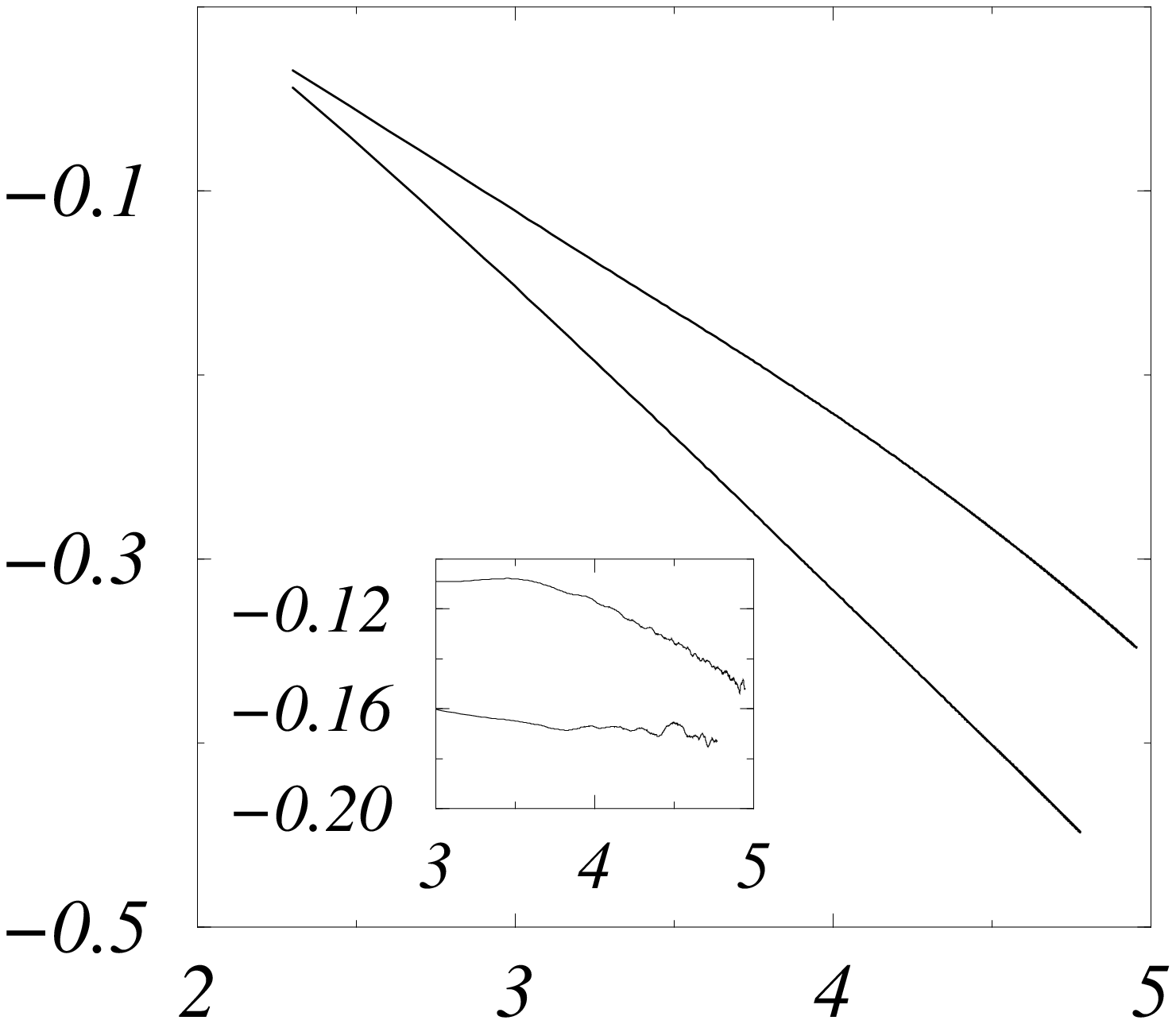}}}
\put(0,185){\makebox(0,0){\large $\log p$}}
\put(180,0){\makebox(0,0){\large $\log t$}}
\put(170,140){\makebox(0,0){$\scriptstyle g=0.172$}}
\put(170,130){\vector(0,-1){20}}
\put(70,110){\makebox(0,0){$\scriptstyle g=0.185$}}
\put(60,120){\vector(1,1){20}}
\end{picture}
\hspace{30\unitlength}
\begin{picture}(200,200)(0,0)
\put(110,210){\makebox(0,0){\large (b)}}
\put(0,0){\makebox(200,200){\epsfxsize=220\unitlength\epsffile{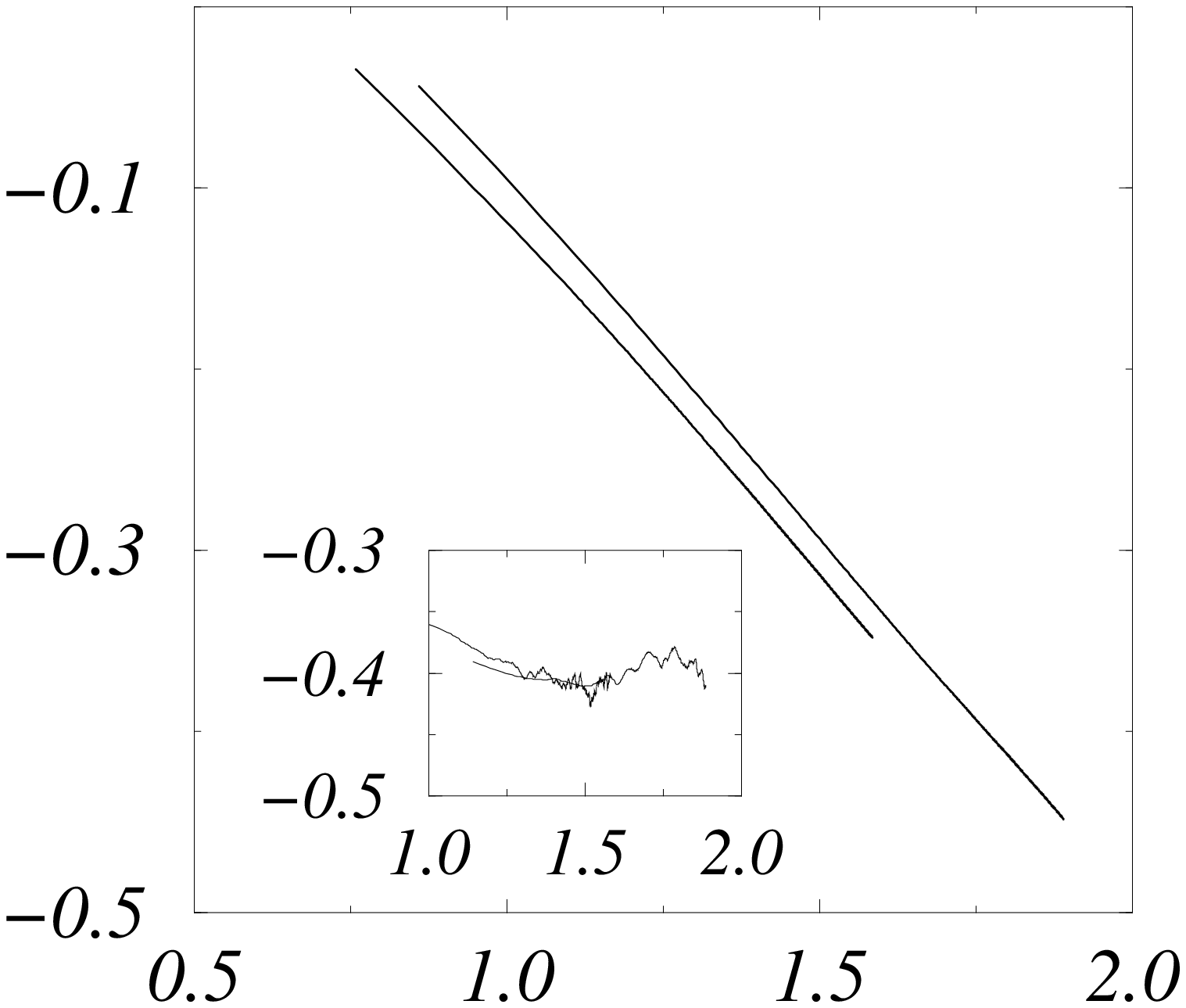}}}
\put(0,185){\makebox(0,0){\large $\log p$}}
\put(180,0){\makebox(0,0){\large $\log L$}}
\put(150,140){\makebox(0,0){$\scriptstyle g=0.185$}}
\put(155,130){\vector(-1,-1){20}}
\put(70,110){\makebox(0,0){$\scriptstyle g=0.172$}}
\put(60,120){\vector(1,1){20}}
\end{picture}
\end{center}
\vspace{10\unitlength}
\caption{From the same runs as those presented in Fig~\ref{f5}:
(a): Decay of the persistent fraction $p(t)$ in log-log scales.
(b): Same, but as a function of $L(t)$, the typical cluster scale.
Inserts: local exponents.}
\label{f6}
\end{figure}

We now turn to the other set of results presented in \cite{lc99}, those related
to the decay of $p(t)$, the fraction of persistent sites. 
Fig.~\ref{f6}a shows the decay of $p(t)$ for the same run as in 
Fig.~\ref{f5}. The behavior of the local exponent clearly changes
at late times, in coincidence with the crossover observed in the growth
of $L(t)$. The persistence exponents estimated in \cite{lc99} thus
reflect  only  the short-time behavior of the coarsening process. 

Given that $L(t)$ reaches its asymptotic behavior rather late,
one would  ideally like to start measuring persistence from an initial
time $t_0$ larger than the crossover time, so that the whole history of
the system coming into account in the decay of $p(t)$ be situated in the
asymptotic scaling regime. Unfortunately, this is basically out of reach of
current computers' power, since it would require very large systems and
their simulation over times much larger than $t_0$. 

On the other hand, it was noticed in \cite{lc99} that the decay of 
the persistence probability recorded as a function of $L(t)$ 
is in a sense ``more universal'', as it seems to be roughly independent
of $g$. Those measurements were performed in the intermediate scaling regime
where the growth of $L(t)$ is not normal.  For later times and larger system
sizes, the variation of $p$ with $L$ is seen to be 
better behaved than the simple decay of $p(t)$ (Fig.~\ref{f6}b). 
The local slope
is approximately constant, and especially so after the crossover time for
the growth of $L(t)$. We also checked that changing $t_0$ does not
 significantly influence the results. We can thus estimate a reliable
persistence exponent, whose variation with $g$ is not significant given our 
numerical accuracy, and which takes approximately the value known for
the TDGL equation, i.e. $\theta \simeq 0.20(2)$.

\subsection{Generalized persistence}

The behavior of persistence seems independent of $g$,
or rather we cannot resolve the possible
variation of exponent $\theta$ as we go from the threshold
(near $g_{\rm e}$) to the continuous-space limit.
Moreover, although the estimated $\theta$ is close to the TDGL value,
the error bar does not allow us to rule out the currently accepted value
for the Ising model.

That $\theta$ is close to the TDGL value is not surprising
if one considers that both our system and the TDGL equation 
are governed by deterministic dynamics.
In the strong coupling limit of CMLs, in particular,
interfaces evolve smoothly
on a continuous space, and their motion is expected to
resemble that of interfaces in TDGL.
On the other hand, near $g_e$, strong lattice/anisotropy effects arise, 
yielding somewhat jittery interface motion, similarly to the discrete 
walls observed in Ising model.

Although persistence seems to depend only weakly on these lattice effects, 
it was recently proposed that these discrepancies become obvious 
when generalized persistence is considered \cite{BCDL}.
In order to define generalized persistence,
let us consider the time-average of the coarse-grained variable 
$\sigma_{\vec r}^t= {\rm sign}(X_{\vec r}^t)$.
At each point $\vec r$ in space and starting from some initial time $t_0$,
it reads,
$$
M_{\vec r}^t= \frac{1}{t-t_0}\sum_{t'=t_0}^t\sigma_{\vec r}^{t'} \; .
$$
Generalized persistence $P(t;x)$ can then be defined as the probability that 
$M_{\vec r}^t$ has always remained above some threshold $x\in[-1,1]$
\cite{PERSIS}:
$$
P(t;x) = {\rm Prob}\left(M^{t'}\ge x;\,\forall t'\le t\right)\,.
$$
Generalized persistence is nothing but the 
``standard'' persistence for the process ${\rm sign}(M^{t}-x)$,
and is thus also expected to decay algebraically in time 
with exponent $\theta(x)$.
This provides a spectrum $\theta(x)$ of generalized persistence
exponents, which contains, in particular, 
the ``standard'' persistence exponent $\theta(1)=\theta$.

As they discriminate in deeper details the possible paths followed by the 
variable $\sigma_{\vec r}^t$, generalized persistence exponents 
are expected to encode more information 
on the properties of interface motion than simple persistence
and, in particular, to be sensitive to the influence of the 
jittery motion of interfaces \cite{BCDL}.

\begin{figure}
\begin{center}
\begin{picture}(350,300)(0,0)
\put(0,0){\makebox(350,300){\epsfxsize=350\unitlength\epsffile{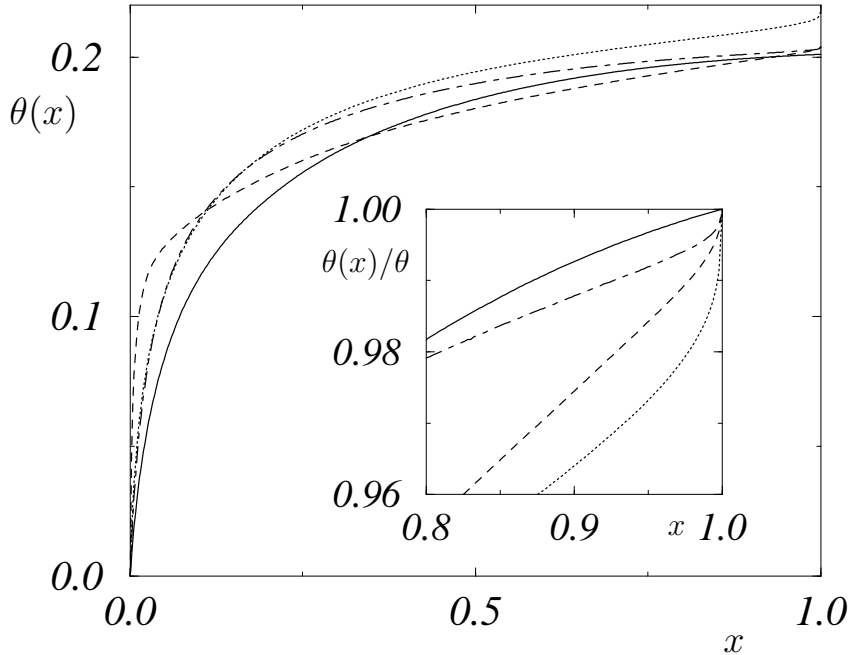}}}
\put(0,240){\makebox(0,0){\large $\theta(x)$}}
\put(300,10){\makebox(0,0){\large $x$}}
\put(140,175){\makebox(0,0){$\theta(x)/\theta$}}
\put(275,60){\makebox(0,0){$x$}}
\end{picture}
\end{center}
\caption{
Generalized persistence exponent spectra $\theta(x)$ for the TDGL equation 
($t_0=100$, $L=8192$, $dt=0.2$, $dt/dx^2=1/8$, $t_{\rm max}\sim 16000$, solid line), 
our CML 
($t_0=100$, $L=8192$, $t_{\rm max}\sim 10^5$) at $g=0.172$ (dashed) and $g=0.185$ (dot-dashed), 
and the zero-temperature Ising model ($t_0=100$, $L=16384$, $t_{\rm max}=25000$, dotted).
Insert: $\theta(x)/\theta$.
}
\label{f7}
\end{figure}

Here we estimate $\theta(x)$ near and away from $g_{\rm e}$.
The measurement of generalized persistence has been carried out on the
same runs as those presented previously.
Spectra of generalized persistence exponents are displayed  in
Fig.~\ref{f7}.
It should first be noted that these spectra are difficult to resolve
numerically for small $x$ values, 
since the corresponding data relies on the relatively rare spins which
have spent most of their time in the phase opposite to their original phase.
Thus, the data presented in Fig.~\ref{f7} for this region
can only have an indicative value (it is believed on general grounds,
however, that $\theta(x)$ goes to zero as $(1+x)^\theta$ when $x\to -1$).

On the other hand, 
the normalized exponents $\theta(x)/\theta$  are rather well-resolved
numerically near $x=1$, and their behavior in this region is
expected to characterize the short-scale motion of interfaces.
The insert in Figure~\ref{f7} shows
$\theta(x)/\theta$ for our CML as well as for the Ising model and the TDGL
equation. The normalized exponent spectra of our CML tend to
the TDGL behavior as $g$ goes away from $g_{\rm e}$ and approaches the
continuous-space limit. 
One can notice, moreover, that our CML exhibits, under
identical experimental conditions, the same value 
$\theta\simeq 0.204$ for both values of $g$ presented here, despite
the important change in the strength of discretisation effects. Thus,
these effects are not felt at the level of simple persistence but are well
captured, in a sense, by generalized persistence. Space discretisation
involves at least two factors: the special behavior of interfaces
(as seen in Section~\ref{interface}) and anisotropy. The results reported 
here do not allow one to separate them but imply that they seem to have 
only a marginal effect on $\theta$. We note, further, that 
this is in agreement with recent studies
of persistence in anisotropic partial differential equations in which 
no detectable change of $\theta$ with anisotropy could be recorded \cite{TBP},
but in disagreement with the claims of \cite{rutenberg}.

\section{Conclusion}

The results presented here have elucidated the surprising coarsening
behavior of chaotic CMLs previously reported in \cite{lc99}, which we
have shown to be representative only of the long, intermediate scaling region
present in such deterministic systems. ``Normal'', $z=2$ phase-ordering
is recovered at large time/length scales for all coupling strengths
$g>g_{\rm e}$. 

In fact, both the growth law and the persistence exponent $\theta$ 
seem to be independent of $g$ in this limit. Although our numerical data
does not allow for a definitive statement about this, it provides an 
interesting case for the current debate considering
which factors ---space-discretization, anisotropy, chaos or
probabilistic motion--- are possibly influencing persistence
exponents \cite{rutenberg,BCDL}.

The general picture emerging for the phase-ordering
properties of our CML is as follows: 
while TDGL-like, smooth interface motion is recovered in
the continuous-space limit,
near $g_{\rm e}$, important lattice/anisotropy
effects yield interfacial properties somewhat
similar to that of the Ising model; but these are only felt in 
generalized persistence scaling, confirming that this quantity captures
details of interface dynamics. 
Generalized persistence spectra $\theta(x)$ 
show a significant qualitative change
as $g$ goes from $g_{\rm e}$ to the continuous-space limit, similar to
the recently observed difference for $\theta(x)$ between the Ising and TDGL.
Thus our CML, to some extent, interpolates between these two models.

In this regard, the ultimate question of the ``true'' asymptotic behavior
of the phase-ordering properties of our CML must depend on further 
knowledge about the status of the commonly
observed numerical differences in  persistence properties between the TDGL
equation and the Ising model. Future work should focus on resolving
the question of observed discrepancy's origin, possibly with the help
of ``intermediate'' systems such as the CML studied here.

\end{document}